\documentclass[a4paper,10pt]{article}
\usepackage[dvips]{color}
\usepackage{amsmath}
\usepackage{amssymb}
\usepackage{graphics}
\usepackage{amscd}

\newtheorem{thm}{Theorem}

\newcommand{\qinv}{q^{-1}}

\newcommand{\sutwo}{$SU(2)$ }

\newcommand{\ie}{\textit{i.e.}}

\newcommand{\eg}{\textit{e.g.}}

\newcommand{\bfl}{\begin{flushleft}}
\newcommand{\efl}{\end{flushleft}}
\newcommand{\bfr}{\begin{flushright}}
\newcommand{\efr}{\end{flushright}}
\begin{document}
\thispagestyle{empty}
\begin{center}
{\Large \bf A $q$-analog of the ADHMN construction}\\ 
\vspace{5 mm}
{\Large \bf and axisymmetric multi-instantons}\\
\vspace{2cm}
Masaru Kamata 
\footnote[1]
{e-mail address: nkamata@minato.kisarazu.ac.jp} \\
  \vspace{0.7cm}
  {\it Kisarazu National College of Technology\\
    2-11-1 Kiyomidai-Higashi, Kisarazu, Chiba 292-0041, Japan}\\
  \vspace{1cm}
Atsushi Nakamula
\footnote[2]
{e-mail address: nakamula@sci.kitasato-u.ac.jp}\\
  \vspace{0.7cm}
{\it Department of Physics, School of Science, Kitasato University\\
Sagamihara 228-8555, Japan}\\
  \vspace{1.5cm}
\end{center}  
%
%
%
%
%
%

\begin{abstract}
In the preceding paper (1999 \textit{Phys.Lett.~}\textbf{B463} 257), the authors presented a $q$-analog of the ADHMN construction and obtained a family of anti-selfdual configurations with a parameter $q$ for classical \sutwo Yang-Mills theory in four-dimensional Euclidean space.
The family of solutions can be seen as a $q$-analog of the single BPS monopole preserving (anti-)selfduality.
Further discussion is made on the relation to axisymmetric ansatz on anti-selfdual equation given by Witten in the late seventies. 
It is found  that the $q$-exponential functions familiar in $q$-analysis appear as analytic functions categorizing the anti-selfdual configurations yielded by axisymmetric ansatz.
\end{abstract}

PACS numbers: 02.30.Gp; 02.30.Jr; 11.15.-q; 11.27.+d
\newpage
\setcounter{page}{2}
\setcounter{section}{0}
\section{Introduction}
The ADHM construction \cite{AHDM,DM} gives (anti-)selfdual ((A)SD) configurations with a finite topological instanton number $k$ for classical $Sp(n)$ Yang-Mills(YM) theories in four-dimensional Euclidean space $\mathbb{R}^4$.
In the construction, an instanton configuration is obtained through a vector space with dimensionality $(n+k)$ \cite{Corri}. 
In ref.\cite{Nahm82} Nahm applied the ADHM construction to the derivation of multi-monopole solutions for $SU(2)$ ($\simeq Sp(1)$) Yang-Mills-Higgs system by introducing an infinite dimensional Hilbert space $\mathcal{L}^2$, following the attempt\cite{Nahm80} to construct the BPS monopole \cite{Bogo,PS}, in which the Hilbert space is  defined on the interval $I:=[-1/2,1/2]$.
The ADHM construction together with Nahm's one are called the ADHMN construction.
In each formulation, it is vital to solve a linear equation $\Delta^\dagger v=0$, where $v$ is a quaternionic vector in appropriate vector space and the operator $\Delta$ is linear in the quaternion spacetime coordinates $x$.
In particular, it turns out to be a differential equation  in Nahm's case.
Further application was made by Braam and Austin \cite{BA}, who gave a discretized version of the Nahm formalism and pointed out the correspondence between the discrete Nahm equation and the hyperbolic monopoles \cite{Ati}, \ie, monopoles on hyperbolic three-space $H^3$.

In ref.\cite{KN} the authors presented another way of the discretization to the Nahm formalism.
Instead of Nahm's infinite dimensional vector space $\mathcal{L}^2$ (with continuous measure), they introduced an infinite dimensional vector space $\ell^2$ (with discrete measure) defined on the ``$q$-interval'' 
$I_q:=\left\{\pm1/2,\pm q/2,\pm q^2/2,\right.$ $\left.\pm q^3/2,\dots\right\}$, where $q \in (0,1)$ is a real free parameter so that the points are multiplicatively placed on the interval $I$.
The interval $I_q$ has an accumulation point at the origin.
In place of a linear algebraic operator in the ADHM construction or a linear differential operator in  Nahm's formulation, we can make use of a linear $q$-difference operator in the $\ell^2$ formulation.
The new formulation is, in other words, a $q$-analog of the ADHMN construction. 
The characteristic is that the discretization scheme is a multiplicative one in the $\ell^2$ formulation in contrast to the hyperbolic monopole case, where the strategy was an additive one.
As a primary application of the $\ell^2$ formulation \cite{KN}, a family of ASD configurations with a parameter $q$ for  \sutwo YM theory in  $\mathbb{R}^4$ was obtained, and it was found that this solution approached the BPS monopole in the limit $q \to 1$.

The aim of our work is to understand the whole structure of the solution space to (A)SDYM theories.
In particular, we have to clarify the position, in the solution space, of the solutions given by the $q$-analog of the ADHMN construction.
For this purpose, we should remind the works done in the late seventies especially by Witten \cite{W} and by Manton \cite{Mant},  about the classical solutions of (A)SDYM without using the ADHMN construction.
We will find in the present paper the one-parameter solution derived by the $q$-ADHMN formalism gives a special interpolation between Witten's axisymmetric multi-instantons and the BPS monopole.
Following Witten, we can classify each axisymmetric multi-instantons by a meromorphic function which has finite number $k$ of poles or zeroes, up to gauge transformation.
Manton \cite{Mant} considered a large-action limit $k \to \infty$ of a multi-instantons configuration, where the meromorphic function  approached the exponential function and then the BPS monopole was derived \cite{Mant}. 
Hence we can see that the position of the BPS monopole in the solution space of (A)SDYM is on the extremity, in a sense.
In comparison with monopoles, we may characterize the one-parameter solution by the $q$-exponential functions, familiar in $q$-analysis \cite{GR}.
In the limit $q \to 1$ the infinite poles of the $q$-exponential function are approaching infinity, where the ordinary exponential function corresponding to the BPS monopole has the essential singularity, the one-parameter solution also approaches the BPS monopole in this limit, accordingly. 

In Sects.2 and 3 we briefly review the ADHM construction and Nahm's one, respectively. 
In Sect.4, we summarize the $q$-analog of the ADHMN construction.
In Sect.5, we review Witten's axisymmetric multi-instantons and Manton's derivation of the BPS monopole as a large-instanton number limit, and discuss the relation between the one-parameter solution and axisymmetric multi-instantons.
Sect.6 is devoted to concluding remarks.
\section{The ADHM construction of $k$-instanton solutions}

In this section, we briefly summarize the ADHM construction \cite{AHDM,DM,Corri}
\footnote[1]{We use the same notation as that of \cite{KN} unless otherwise stated. For example, symbol $\dagger$ denotes hermitian conjugation, $\tau_\mu=(1_2,i\sigma_1,i\sigma_2,i\sigma_3)$ and $x_\mu=(x_0,x_1,x_2,x_3)$ are quaternion elements and spacetime coordinates, respectively, $x:=\sum^3_{\mu=0}x_\mu\tau_\mu$, $|x|^2:=\sum^3_{\mu=0}x_\mu x_\mu$, and $ \hat x:=\sum_{j=1}^3x_j\sigma_j/r$.}.
The Euler-Lagrange equations, $D_\mu F_{\mu\nu}=0$, in classical $Sp(n)$ Yang-Mills theories in $\mathbb{R}^4$ are automatically satisfied by the (A)SD equations $F_{\mu\nu}=\pm\tilde F_{\mu\nu}$ due to the Bianchi identity $D_\mu \tilde F_{\mu\lambda}=0$.
The ADHM construction gives (A)SD configurations with a finite instanton number $k$, which are obtained through a vector $v$ of an $(n+k)$-dimensional quaternion vector space $V^{n+k}$ with inner product $<w,v>:=w^\dagger v$.
The connection one-form is given by
\begin{equation}
A(x)=i<v, dv>=iv^\dagger(x)dv(x) \label{eq:A}
\end{equation}
and is (A)SD due to the theorem:
\begin{thm} {\rm \cite{AHDM}} 
For $Sp(n)(\supset U(n),\;O(n))$ gauge group, the $(n+k)\times n$ matrix $v$ enjoying a linear equation 
$\Delta^\dagger v=0$ and normalization $v^\dagger v=1_n$ yield (A)SD gauge fields, if the matrix $\Delta^\dagger \Delta$ is quaternionic real and invertible. Here the $(n+k)\times k$ matrix $\Delta$ is assumed to be linear in $x$, \ie, $ \Delta=a+bx$.
\end{thm}

We can trace the proof of this theorem in the following way.
First of all, we notice that the $n+k$ column vectors of the matrices $v$ and $\Delta$ span $V^{n+k}$, which can be understood from the normalization $v^\dagger v=1_n$, the invertibility of the matrix $\Delta^\dagger \Delta$,
and the linear equation  $\Delta^\dagger v=<\Delta,v>=0$ which implies that the column vectors of $v$ and  those of $\Delta$ are orthogonal to each other.
Then we can find the completeness condition, $1_{n+k}=v(v^{\dagger}v)^{-1}v^{\dagger}+\Delta(\Delta^{\dagger}\Delta)^{-1}\Delta^{\dagger}$.
This gives the following two projection operators,
$P:=v(v^{\dagger}v)^{-1}v^\dagger=vv^\dagger$ and $P':=\Delta(\Delta^{\dagger}\Delta)^{-1}\Delta^{\dagger}=\Delta \mathcal{F} \Delta^{\dagger}$, where $\mathcal{F}:=(\Delta^{\dagger}\Delta)^{-1}$, onto the $n$ and $k$ dimensional subspaces spanned by the column vectors of $v$ and $\Delta$, respectively.
We obtain the curvature two-form from (\ref{eq:A})  expressed in terms of $P$ and $v$ as
\begin{eqnarray}
F&=&dA-iA\wedge A\nonumber\\
&=&iv^{\dagger}dP \wedge dP\,v\nonumber\\
&=&iv^{\dagger}b\, dx\,\mathcal{F}\wedge dx^{\dagger}\,b^{\dagger}v,
\end{eqnarray}
where we have used the assumption $\Delta=a+bx$. 
Since $\Delta^\dagger \Delta$ is quaternionic real, \ie, each entry is proportional to $1_2$, $\mathcal{F}$ is also quaternionic real, hence commutes with the quaternion coordinate $x$.
We find that $F$ is ASD, because
\begin{equation}
dx \wedge dx^{\dagger}=i\bar{\eta}^j_{\mu\nu}\sigma_jdx_\mu 
\wedge dx_\nu,
\end{equation}
where $\bar{\eta}^j_{\mu\nu}$ is the 't Hooft ASD tensor \cite{'tH}.
On the other hand, a SD curvature two-form is also derived by exchanging $x$ and $x^\dagger$, which yields the 't Hooft SD tensor ${\eta}^j_{\mu\nu}$ \cite{'tH}
\begin{equation}
dx^{\dagger} \wedge 
dx=i\eta^j_{\mu\nu}\sigma_jdx_\mu \wedge dx_\nu.
\end{equation}

Note that the gauge group acts on $v$ on the right, $v \to v'=vg$ where $g \in Sp(n)$, then the connection one-form transforms correctly, $A \to A'=g^\dagger A g+g^\dagger idg$.
And we can prove that the integral $(16\pi^2)^{-1}\int\;{\rm tr}(F \wedge F)$ gives the instanton number $k$.

To be concrete, we hereafter restrict ourselves to the gauge group  $SU(2) \simeq Sp(1)$, that is, $n=1$.
In this case the matrix $v$ reduces to a $(1+k) \times 1$ quaternionic matrix, \ie, a column vector of $(1+k)$ quaternionic components.
For the $k=1$ case, we can construct a one-instanton solution, known as the BPST solution \cite{BPST}.
We set the matrix $\Delta$ as follows without loss of generality, 
\begin{equation}
\Delta =\left(
        \begin{array}{@{\,}cc@{\,}}
        x-i\lambda 1_2 \\
        x+i\lambda 1_2
        \end{array}
        \right).
\end{equation}
This leads to the well known connection one-form in regular gauge,
\begin{equation}
A=\bar{\eta}^j_{\mu\nu}\frac{\sigma_jx_\nu}{|x|^2+\lambda^2}dx_\mu.
\end{equation}
For $k>1$ cases, the canonical form \cite{AHDM,EWein} of $\Delta$ is known, that is,
\begin{equation}
\Delta =\left(
        \begin{array}{@{\,}cccc@{\,}}
        \lambda_1  1_2 & \lambda_2  1_2 & \cdots & \lambda_k 1_2  \\
        x+\alpha_1 1_2 & 0              & \cdots & 0              \\
        0              & x+\alpha_2 1_2 & \cdots & 0              \\
        \cdots         & \cdots         & \cdots & \cdots         \\
        0              & 0              & \cdots & x+\alpha_k 1_2 \\
        \end{array}
        \right).
\end{equation}
These yield the 't Hooft singular gauge multi-instanton solutions with $5k-3$  effective parameters \cite{AHDM}
\begin{equation}
A=\frac{1}{2}{\eta}^j_{\mu\nu}\sigma_j\partial_\mu \ln \Biggl(1+\sum^k_{i=1}\frac{\lambda^2_i}{|x+\alpha_i|^2}\Biggr)dx_\nu. 
\end{equation}

The most general multi-instantons solutions admit $8k-3\;(k>1)$  effective parameters \cite{AHDM,DM}, however any explicit formula for these solutions has never been known.

\section{Nahm's derivation of monopoles}
In static Yang-Mills-Higgs system, the Bogomol'nyi equations \cite{Bogo}, which governs minimum energy configurations, are $D_j\Phi=\pm \frac{1}{2}\epsilon_{jkl} F_{kl}$.
This system can be regarded as static YM if we identify $\Phi=A_0$.
As mentioned in introduction, Nahm \cite{Nahm82} applied the ADHM construction to constructing monopoles, localized field configurations in $\mathbb{R}^3$ to the Bogomol'nyi equations, by bringing in an infinite dimensional vector space $\mathcal{L}^2$.
Intuitively, we can recognize the necessity of the infinite dimensional $\mathcal{L}^2$ space through the following argument \cite{Rossi}.
We consider a single monopole as a superposition of instantons putting densely on a time axis at a definite location in $\mathbb{R}^3$.
To compose this configuration we firstly put the instantons periodically on each time axis, this situation is called caloron solution \cite{KvB,HS}, and then we take the limit of infinitesimal periodicity to restore time translation invariance.
Obviously, the instanton number of monopoles is infinite, so that we need an infinite dimensional vector space in the language of the ADHM construction.

In Nahm's construction, we assume a linear equation $\Delta^\dagger v=0$ for the vector $v(z) \in \mathcal{L}^2[I]\otimes V_N \otimes \mathbb{H}$ which defines monopole configurations, where $V_N$ is an additional $N$-dimensional vector space representing a multi-monopoles configuration.
The matrices $\Delta$ and $\Delta^\dagger$ of the ADHM construction turn out to be differential operators here.
The connection one-form is given by the formula (\ref{eq:A}) with an $\mathcal{L}^2$ inner product,  
\begin{equation}
<w,v>=\int_I w(z)^\dagger v(z)\,dz. 
\end{equation}
We find that the conditions on $\mathcal{F}$ become,
\begin{thm}{\rm \cite{Hit83}, \cite{Nahm82}}
If
\begin{equation}
\Delta^\dagger=i\frac{d}{dz}\otimes1_N\otimes1_2+1\otimes1_N\otimes x^\dagger+\sum_{j=1}^31\otimes iT_j(z)\otimes\tau_j^\dagger,
\end{equation}
 the quaternionic reality and invertibility of $\Delta^\dagger\Delta$ are equivalent to the differential equations for the matrices $T_j$ (the Nahm equation),
\begin{equation}
\frac{dT_j}{dz}=\frac{1}{2}\epsilon_{jkl}[T_k,T_l]. \label{Nahmeq}
\end{equation}
\end{thm}
In practice we need some additional conditions on $T_j$'s  to impose the correct boundary conditions to monopoles, \eg, $\Phi=1+k/r+O(r^{-2})$ as $r\to\infty$, where $k$ is the quantity representing magnetic charge.

The simplest example of solutions to the Nahm equation (\ref{Nahmeq}) is $T_j(z)=0 \,\, (j=1,2,3)$ \cite{Nahm82}, which leads to the single BPS monopole.
 The linear equation $\Delta^\dagger v=0$ with $\Delta^\dagger=i\frac{d}{dz}+x^\dagger$ gives $v(z)=N(x_\mu)e^{ix^\dagger z}$, where $N(x_\mu)$ is a normalization function, hence we arrive at the following connection one-form of the BPS monopole
\begin{equation}
A_{BPS}=-\frac{1}{2}\Bigl(\coth r-\frac{1}{r}\Bigr)\hat{x}dt-\frac{1}{2}\Bigl(1-\frac{r}{\sinh r}\Bigr)\epsilon_{ijk}\frac{x_i}{r^2}dx_j\sigma_k\label{A_BPS}
\end{equation}

\section{A  $q$-analog of the ADHMN construction -- introducing $\ell^2$ vector space}
So far, we have seen the outline of the ADHMN construction for (A)SD configuration especially in \sutwo case, and found that the dimensionality of the vector space $V$ with which each solution is associated is crucial to determine the instanton number of field configurations.
In particular, if we needed a monopole solution by this formalism, we had to introduce an infinite dimensional vector space.

In ref.\cite{KN} the authors showed that there existed another way of introducing $\mbox{dim}(V)=\infty$.
Instead of Nahm's infinite dimensional vector space $\mathcal{L}^2$ with continuous measure, they introduced an infinite dimensional vector space with discrete measure, the $\ell^2$ vector space.
The $\ell^2$ space is the function space defined on the $q$-interval, $I_q=\left\{\pm1/2,\pm q/2,\pm q^2/2,\pm q^3/2,\dots\right\}$, where $q \in (0,1)$ is a real free parameter so that the points are multiplicatively placed in the region $I=[-1/2,1/2]$.
Notice that the interval $I_q$ has an accumulation point at the origin. 
In place of a linear differential operator $\Delta$ in monopole construction, we must introduce a linear $q$-difference operator in the $\ell^2$ formulation.
This is the reason why we call the $\ell^2$ formulation a $q$-analog of the ADHMN construction, or $q$-ADHMN in short.
Here we sketch out the $\ell^2$ formulation and find that we can produce, as the simplest example, a family of ASD configurations with a parameter $q$ for \sutwo YM theory in $\mathbb{R}^4$.

We begin with defining the appropriate inner product $<w,v>_q$ of the $\ell^2$ vector space on $I_q$ as
\begin{equation}
<w,v>_q\,=\int^{1/2}_{-1/2}w^* v \,d_qz:= \int^{1/2}_0w^* v \,d_qz-\int^{-1/2}_0w^* v \,d_qz, \label{<,>_q}
\end{equation}
where the integration is defined by the $q$-integral (``Thomae-Jackson integral") \cite{GR}, in fact it is an infinite summation,
\begin{equation}
\int^{a}_0f(z)\,d_qz:=a(1-q)\sum^\infty_{n=0}f(aq^n)q^n. \label{DefTJ}
\end{equation} 
In (\ref{<,>_q}) the conjugate vector $v^*$ is defined as
\begin{equation}
v^*=[v(x_\mu,z;q)]^*\ :=\ v^\dagger(x_\mu,qz;\qinv), \label{*}
\end{equation}
where $\dagger$ is hermitian conjugation.
Note that the asterisk is an involution: $(v^{*})^{*}=v$.
The inner product is crucial to yield (A)SD configurations through the connection one-form
\begin{equation}
A=i<v,dv>_q.\
\end{equation}

We demonstrate that the $\ell^2$ formulation actually works to produce an ASD configuration in its simplest case, a SD one is similarly constructed.
To do this, analogously to the BPS monopole construction given in the last section, we consider a linear $q$-difference equation which determines the vector $v$,  $\Delta^* v=0$,  where
\begin{equation}
\Delta^*=iD_q+x^\dagger. \label{Dast}
\end{equation}
Here the $q$-derivative $D_q$ is defined as
\begin{equation}
D_qf(z):=\frac{f(z)-f(qz)}{(1-q)z}. 
\end{equation}
Since the relation
\begin{equation}
 <iD_qw,v>_q\,=\,<w,iD_qv>_q-i\Bigl[w^\dagger(x_\mu,z;q^{-1})v(x_\mu,z;q)\Bigr]^{z=1/2}_{z=-1/2}, \label{selfadj}
\end{equation}
holds, we can prove the self-adjointness for the operator $iD_q$ when we consider a certain class of $\ell^2$ functions.
Then we have
\begin{equation}
\Delta=iD_q+x 
\end{equation}
due to the self-adjointness of $iD_q$, this leads to the quaternionic reality of the product $\Delta^*\Delta$, that is,
\begin{eqnarray}
\Delta^*\Delta&=&-D_q^2+2itD_q+|x|^2\nonumber\\
     &=&(iD_q+\rho_{+})(iD_q+\rho_{-})1_2,
\end{eqnarray}
where $\rho_\pm:=t\pm ir=x_0\pm ir$ and we have explicitly shown the real part of quaternion $1_2$ in the last line.
We can show \cite{KN} that there exists a function satisfying the equation,
\begin{equation}
\Delta^*\Delta \mathcal{F}(x_\mu;z,z';q)=\frac{1}{(1-q)|z'|}\delta_{z,z'},
\end{equation}
which implies the invertibility of $\Delta^*\Delta$.
If we consider the limit $q \to 1$, then $I_q \to I$ and $D_qf(z) \to \frac{df(z)}{dz}$, thus the formulation turns out to be that of the BPS monopole, in fact the function $\mathcal{F}$ approaches Nahm's one \cite{Nahm80}
\begin{equation}
\mathcal{F} \xrightarrow [q\to1]{}-\frac{1}{2r}e^{it(z-z')}\sinh r|z-z'|+\mathcal{F}_0,
\end{equation}
where $\mathcal{F}_0$ is a kernel of $\Delta^*\Delta$ to be determined by boundary conditions, which does not need fixing here.
We therefore confirm that the linear operator (\ref{Dast}) under the inner product (\ref{<,>_q}) makes the curvature ASD.

Now we solve the linear equation $\Delta^*v=0$ for the vector $v$ defined on $I_q$.
For this purpose we introduce ``$q$-exponential functions"
\begin{eqnarray}
e_q(z)&=&\prod^{\infty}_{n=0}(1-q^nz)^{-1}=\sum^\infty_{n=0}\frac{z^n}{(q;q)_n}, \quad(|z|<1),\\
E_q(z)&=&\prod^{\infty}_{n=0}(1+q^nz)=\sum^\infty_{n=0}\frac{q^{\frac{n(n-1)}{2}}}{(q;q)_n}z^n,
\end{eqnarray}
which obey the relation $e_q(z)E_q(-z)=1$. 
Then we can easily find that the solution to the linear equation is 
\begin{equation}
v=e_q(ix^\dagger (1-q)z)N(x_\mu;q),\label{v=eN}
\end{equation}
where $N(x_\mu;q)$ is a ``normalization function", putting on the right of $e_q$ to avoid the ordering ambiguity due to quaternion calculus.
We obtain the normalization function \footnote[1]{The functional form of $N$ will have ambiguity, see Sect.6.}
\begin{eqnarray}
N(x_\mu;q)&=\frac{1}{2}\{(\Lambda_++\Lambda_-)^{-\frac{1}{2}}+(\Lambda_+-\Lambda_-)^{-\frac{1}{2}}\}&1_2\nonumber\\
&+\frac{1}{2}\{(\Lambda_++\Lambda_-)^{-\frac{1}{2}}-(\Lambda_+-\Lambda_-)^{-\frac{1}{2}}\}&\hat x, \label{N}
\end{eqnarray}
where $\Lambda_{\pm}$ are defined by
\begin{eqnarray}
&&<e_q(ix^\dagger (1-q)z),e_q(ix^\dagger (1-q)z)>_q \nonumber\\
&&=\frac{\hat x}{2r}\Bigl[E_q(-ix(1-q)z)e_q(ix^\dagger (1-q)z)\Bigr]^{1/2}_{z=-1/2}\nonumber\\
&&=\Lambda_+(t,r;q)1_2+\Lambda_-(t,r;q)\hat x,\label{e,e}
\end{eqnarray}
with\footnote{The expression $(\frac{\rho_+}{\rho_-};q)_{2n}$ in Eq.(40) of Ref.[10] should read $(q\frac{\rho_+}{\rho_-};q)_{2n}$.}  
\begin{equation}
\Lambda_\pm(t,r;q)
:=\frac{1-q}{2}\left\{\sum_{n=0}^\infty\frac{(q\frac{\rho_+}{\rho_-};q)_{2n}}{(q;q)_{2n+1}}\biggl(i\frac{(1-q)\rho_-}{2}\biggr)^{2n}\pm (\rho_+\leftrightarrow\rho_-)\right\}.\label{Apm}
\end{equation}

The connection one-form  given by (\ref{v=eN}) with (\ref{N}) is expressed in the formula
\begin{equation}
A=\frac{1}{4}\Bigl(-\frac{\partial\Omega}{\partial r}dt+\frac{\partial\Omega}{\partial t}dr\Bigr)+f d\hat{x}+g\epsilon_{ijk}\frac{x_i}{r^2}dx_j\sigma_k, \label{qbps}
\end{equation}
where
\begin{eqnarray}
\Omega(x_\mu)&:=&\log\frac{L^+}{L^-}\cdot 1_2+\log(L^+L^-)\cdot\hat{x},\label{Omega}\\
f(t,r)&:=&-\frac{i}{4}(M_+-M_-)(L^+L^-)^{-1/2},\label{f}\\
g(t,r)&:=&-\frac{1}{2}\Bigl\{1-\frac{M_++M_-}{2}(L^+L^-)^{-1/2}\Bigr\}\label{g}.
\end{eqnarray}
Here $L^\pm$ and $M_\pm$ are the following functions of $\rho_\pm=x_0\pm ir$ 
\begin{eqnarray}
L^\pm:=\Lambda_+\pm\Lambda_-=\sum^\infty_{n=0}\frac{(q\frac{\rho_\mp}{\rho_\pm};q)_{2n}}{(q^2;q)_{2n}}\left\{-\frac{(1-q)^2\rho_\pm^2}{4}\right\}^n, \label{Lpm}\\
M_\pm:=\sum^\infty_{n=0}\frac{1-q}{1-q^{2n+1}}\left\{-\frac{(1-q)^2\rho_\pm^2}{4}\right\}^n. \label{Mpm}
\end{eqnarray}
In the limit $q \to 1$ we see that the solution (\ref{qbps}) approaches the BPS one $A_{BPS}$ of (\ref{A_BPS}) as expected
\footnote[2]{We have $A^*=A$ instead of the hermiticity $A^\dagger=A.$
 However, we can obtain $su(2)$-valued connection through a ``gauge transformation" with $g^*=g^{-1}$ \cite{KN}.}, because in this limit
\begin{equation}
L^{\pm} \rightarrow \frac{\sinh{r}}{r}, \quad M_{\pm} \rightarrow 1.
\end{equation}

In the other limit $q \to 0$, we observe the connection approaches pure gauge, \ie, $F\to 0$.

\section{Axisymmetric multi-instantons and the $\ell^2$ construction}
\subsection{Witten's ansatz and the BPS monopole as degenerate instantons}

As mentioned in introduction, Witten gave a systematic way to generate special  solutions to \sutwo (A)SD equations in his early work \cite{W}.
This construction is independent of the ADHMN reviewed in the previous sections, thus we are interested in the connection between them.
In his construction, a component of the (A)SD equations is reduced to the Liouville equation through the spherically symmetric ansatz in $\mathbb{R}^3$, thus we can find special multi-instanton configurations by taking appropriate assumptions on the general solution to the Liouville equation.
We can consider the instantons depend only on $(t,r)$ ``upper" half-plane, so call them ``axisymmetric" multi-instantons.
Particular interest is concerning with the large instanton number limit, \ie, according to Manton \cite{Mant}, the axisymmetric multi-instantons are getting close to the BPS monopole.
In this section we see the relationship between instantons, monopoles and the ``$q$-monopole" constructed in the last section, in the context of Witten's method.

Supposing that the field configuration is spherically symmetric in $\mathbb{R}^3$ we can reduce the four-dimensional (A)SDYM system into an Abelian-Higgs system in a two-dimensional constant negative curvature space being represented by $(t,r)$ upper half-plane.
Here we consider ASD system, SD system can be obtained similarly.
The ansatz on the connection components is,
\begin{eqnarray}
A^a_0&=&-A_0\frac{x_a}{r},\label{WA0}\\
A^a_i&=&-A_1\frac{x_ix_a}{r^2}-\frac{\phi_1}{r^3}(\delta_{ia}r^2-x_ix_a)-\frac{\phi_2+1}{r^2}\epsilon_{iak}x_k,\label{WA1}
\end{eqnarray}
where $A_0, A_1, \phi_1$ and $\phi_2$ are functions of $t$ and $r$.
The ASD equations are reduced to
\begin{eqnarray}
\partial_0\phi_1+A_0\phi_2&=&-\partial_1\phi_2+A_1\phi_1\label{preCR1},\\
\partial_1\phi_1+A_1\phi_2&=&\partial_0\phi_2-A_0\phi_1\label{preCR2},\\
r^2(\partial_0A_1-\partial_1A_0)&=&-(1-\phi_1^2-\phi_2^2)\label{preLiouville},
\end{eqnarray}
where $\partial_0=\partial/\partial t$ and $\partial_1=\partial/\partial r$, respectively.
Setting an appropriate choice of gauge condition $\partial_\mu A_\mu=0$ so that $A_\mu=-\epsilon_{\mu\nu}\partial_\nu\psi$ for a function $\psi$, we find that (\ref{preCR1}) and (\ref{preCR2}) are the Cauchy-Riemann equations for a complex function $f=e^{-\psi}\phi_1+ie^{-\psi}\phi_2$.
By using complex coordinates $\rho_\pm=t\pm ir$, the remaining component of ASD equation (\ref{preLiouville}) turns out to be the Liouville equation,
\begin{equation}
\partial_{+}\partial_{-}\phi=2e^{-\phi} \label{Liouville}
\end{equation}
where $\phi(\rho_+,\rho_-)$ is a function determined by $\phi_j$ and $\psi$.
The general solution to the Liouville equation (\ref{Liouville}) is well known to be, in terms of arbitrary analytic functions $g_+(\rho_+)$ and $g_-(\rho_-)$, 
\begin{equation}
\phi=-\ln \frac{g'_{+}g'_{-}}{(g_{+}-g_{-})^2}. \label{solLiouville}
\end{equation}
In particular, if $g_+$ is a meromorphic function with $k$ zeroes and poles,
\begin{equation}
g_+=\prod_{i=1}^k\frac{a_i-\rho_+}{\bar a_i+\rho_+}\label{meromfunc},
\end{equation}
for $a_i$ some complex numbers, then the gauge field configuration has finite instanton number $k-1$.

Our present interest is the large $k$ limit of (\ref{meromfunc}), since it corresponds to the $\mathcal{L}^2$ formulation reviewed in Sect.3 or $\ell^2$ formulation reviewed in Sect.4, of ADHMN construction.
The infinite $k$ case of axisymmetric multi-instantons was considered firstly by Manton \cite{Mant}, and showed that the field configuration was gauge equivalent to the BPS monopole.
To trace the construction, we choose the meromorphic function in which all the zeroes and poles are degenerated,
\begin{equation}
g^{(k)}_+(\rho_+)=\Biggl(\frac{1-{i\beta \rho_+}/{2k}}{1+{i\beta \rho_+}/{2k}}\Biggr)^k, \label{BPSmero}
\end{equation}
and take the limit $k\to\infty$.
In this limit, the degenerate zeroes and poles in (\ref{BPSmero}) are approaching infinity and composing an essential singularity, simultaneously.
In fact, (\ref{BPSmero}) turns out to be the exponential function, $\lim g^{(k)}_+=\exp(-i\beta \rho_+)$, and we can obtain the BPS monopole \cite{Mant}.

We therefore find out, in the axisymmetric cases, that  multi-instantons are characterized by a meromorphic function whose singularities are only poles, on the other hand the BPS monopole is described by exponential function, a function with essential singularity.
In terms of the ADHMN construction, we can interpret that the appearance of essential singularity in monopole is a result of Nahm's $\mathcal{L}^2$ formulation.
To explain this, we make an analysis of the solution obtained by the $\ell^2$, \ie, $q$-ADHMN, construction in comparison with monopole, in the next subsection. 
\subsection{The relation to $\ell^2$ construction -- $q$-exponential function as an extremity of meromorphic function}
One can easily see that the connection one-form (\ref{qbps}), together with (\ref{Omega})-(\ref{Mpm}), is in the axisymmetric form (\ref{WA0}) and (\ref{WA1}).
Here we show that the connection is actually given by certain meromorphic functions.
To see this, we examine the component $F_{tr}=-\tilde F_{tr}$, \ie,
\begin{equation}
\partial_+\partial_- \Omega=\frac{1}{2r^2}\Bigl(1-\frac{M_+M_-}{L^+L^-}\Bigr)\hat{x},
\end{equation}
which can be divided into two parts, the trace and traceless part,
\begin{eqnarray}
\partial_+\partial_- \ln \Bigl(\frac{L^+}{L^-}\Bigr)&=&0,\label{Heine} \\ 
\partial_+\partial_- \ln (L^+L^-)&=&\frac{1}{2r^2}\Bigl(1-\frac{M_+M_-}{L^+L^-}\Bigr).
\label{Liouville(a)}
\end{eqnarray}
Although the functions $L^\pm$ are not (anti-)analytic ones, (\ref{Heine}) identically holds since the ratio of the functions $L^+/L^-$ is expressed as the ratio of analytic and anti-analytic functions,
\begin{equation}
\frac{L^+}{L^-}=\frac{(-\kappa^2\rho_+^2;q)_{\infty}}{(-\kappa^2\rho_-^2;q)_{\infty}},
\end{equation}
due to  Heine's transformation formula for $_2\phi_1$ \cite{GR}.
For the rest part (\ref{Liouville(a)}), we can manipulate the infinite series in the product $L^+L^-$ into the combination of infinite product,
\begin{equation}
L^+L^-=\frac{(p_-m_+-p_+m_-)^2}{p_+m_+p_-m_-}\frac{-1}{(\rho_+-\rho_-)^2},
\end{equation}
where 
\begin{eqnarray}
p_\pm(\rho_\pm)&=&\left(i\frac{1-q}{2}\rho_\pm;q\right)_\infty,\\
m_\pm(\rho_\pm)&=&\left(-i\frac{1-q}{2}\rho_\pm;q\right)_\infty.
\end{eqnarray}
We easily find that (\ref{Liouville(a)}) is in the form of the Liouville equation (\ref{Liouville}), with 
\begin{equation}
\phi=-\ln\frac{\left(i\frac{m_+}{p_+}M_+\right)\left(i\frac{m_-}{p_-}M_-\right)}{\left(\frac{m_+}{p_+}-\frac{m_-}{p_-}\right)^2}.
\end{equation}
Finally, we see that the other components of ASD equations are equivalent to the Cauchy-Riemann equation.

Now we observe that the meromorphic function $g_+$ (similarly for $g_-$) appeared in (\ref{solLiouville}) for the  solution under consideration is,
\begin{equation}
g_+=\frac{m_+}{p_+}=\prod_{n=0}^\infty\frac{\frac{2iq^{-n}}{1-q}+\rho_+}{\frac{2iq^{-n}}{1-q}-\rho_+},\label{meromorphicg}
\end{equation}
and its derivative complies with $g_+'=im_+ M_+/p_+$.
The meromorphic function (\ref{meromorphicg}) has infinite zeroes and poles which are not degenerated, and can be read in terms of $q$-exponential functions as $g_+=e_q(-(1-q)i\rho_+/2)E_q(-(1-q)i\rho_+/2)$, so that approaches the ordinary exponential function $g_+ \to e^{-i\rho_+}$ as $q$ tends to $1$.
By taking the limit, all the independent zeroes and poles in (\ref{meromorphicg}) degenerate into one essential singularity at the infinity, then we retrieve the BPS monopole constructed by Nahm's $\mathcal{L}^2$ formulation.

As we saw, in the axisymmetric ansatz, we can classify the solutions into three types by the form of analytic function $g_+(\rho_+)$ (and also $g_-$).
Firstly, a meromorphic function with $k<\infty$ zeroes and poles is characterizing a multi-instanton.
Next, the solution (\ref{qbps}) obtained by the $\ell^2$ construction is identified by a meromorphic function with infinite number of zeroes and poles, a $q$-exponential function.
And finally, the function with essential singularity, the ordinary exponential function leads to a monopole, which can be constructed by $\mathcal{L}^2$ formulation.

\section{Concluding remarks}
As pointed out in Sect.4, we can not determine uniquely the solution (\ref{v=eN}) to the linear $q$-difference equation $\Delta^*v=(iD_q+x^\dagger )v=0$, if we consider the function (\ref{v=eN}) on the continuous interval $I=[-1/2,1/2]$ rather than the discrete point set $I_q$. 
That is, if there exists a function with the following property:
\begin{equation}
C(qz)=C(z), \label{pconst}
\end{equation}
on the continuous interval $I$, being called ``pseudo constant", then we have another solution,   
\begin{equation}
v=C(z)e_q(ix^\dagger (1-q)z)N'(x_\mu;q).\label{v=CeN} 
\end{equation}
The well known pseudo constants are given by Jacobi's theta function $\Theta(z)$ \cite{FR}, for example,
\begin{equation}
C(z)=z^{\alpha}\frac{\Theta(q^{\alpha}z)}{\Theta(z)}, \label{Jtheta} 
\end{equation}
where $\alpha$ is a constant.
The $\ell^2$ formulation discussed in this paper does not have this sort of ambiguity because a function with the property (\ref{pconst}) on $I_q$ is only a constant.
However, we can also successfully formulate the $q$-ADHMN construction with a continuous interval $I$, where the inner product should be defined by ordinary integral, so that the ambiguity occurs in this case.
We will discuss this subject elsewhere.

We can also reformulate the Nahm equation (\ref{Nahmeq}) on the interval $I_q$.
Namely, the (A)SD condition leads to the ``$q$-Nahm equation" in the form   
\begin{equation}
D_qT_i(z)=\frac{1}{2}\epsilon_{ijk}(T_j(qz)T_k(z)-T_k(qz)T_j(z)).
\end{equation}
Although it is not yet understood which boundary conditions for the connection one-form must satisfy, it will be a future work to fix the correct boundary conditions for the matrices $T_j$'s here, similarly to the monopole constructions.
It will be an intriguing topic in integrable systems to consider the interrelation between the ordinary Nahm equation, the discrete Nahm equation and this $q$-difference equation.


\end{document}